\documentclass[5p]{elsarticle}
\makeatletter
\def\ps@pprintTitle{%
 \let\@oddhead\@empty
 \let\@evenhead\@empty
 \def\@oddfoot{}%
 \let\@evenfoot\@oddfoot}
\makeatother

\usepackage{csquotes}
\usepackage{graphicx,color}
\usepackage{float}
\usepackage[caption=false]{subfig}
\usepackage{amsmath}
\usepackage{amssymb}
\usepackage{multirow}
\usepackage{dsfont}
\usepackage{adjustbox}
\usepackage{pdflscape}
\usepackage{wasysym}

\begin{document}

\title{Wall cratering upon high velocity normal dust impact\vspace*{-3.0mm}}
\author{Panagiotis Tolias$^{a}$, Marco De Angeli$^{b}$, Dario Ripamonti$^{c}$, Svetlana Ratynskaia$^{a}$, Giulio Riva$^{c}$, Giambattista Daminelli$^{c}$ and Monica De Angeli$^{b}$}
\address{$^a$Space and Plasma Physics - KTH Royal Institute of Technology, Teknikringen 31, 10044 Stockholm, Sweden\\
         $^b$Institute for Plasma Science and Technology, CNR, via Cozzi 53, 20125 Milano, Italy\\
         $^c$Institute of Condensed Matter Chemistry and Energy Technologies, CNR, via Cozzi 53, 20125 Milano, Italy\vspace*{-12.0mm}}
\begin{abstract}
\noindent Dust-wall high speed impacts, triggered by the termination of runaway electrons on plasma facing components, constitute a source of erosion. Normal high velocity mechanical impacts of tungsten dust on bulk tungsten plates are reproduced in a controlled manner by light gas gun shooting systems. Post-mortem surface analysis revealed that three erosion regimes are realized; plastic deformation, bonding and partial disintegration. The large impact statistics allowed the extraction of reliable empirical damage laws in the latter regime, which can be employed for erosion estimates in future reactors.
\end{abstract}
\maketitle

\section{Introduction}

\noindent The presence of dust in fusion devices has important operational and safety implications, especially in future fusion reactors where metallic dust generation could scale up by orders of magnitude compared to existing devices and nuclear licensing requirements impose several restrictions on the total and hot dust inventories\,\cite{introdu1,introdu2,introdu3}. Hence, the understanding and modelling of dust transport, production, re-mobilization, adhesion and survivability have received wide attention within the fusion community\cite{introdu4,introdu5,introdu6,introdu7,introdu8,introdu9}. Dust-wall mechanical impacts have emerged as an essential feature of dust dynamics in tokamaks\,\cite{dustimp1}. As a consequence of the curved ion flow and inertial effects, mechanical impacts, that span a very wide range of incident speeds, are unavoidable. They have been revealed to control dust migration in the fusion boundary plasma\,\cite{dustimp1}, to influence dust survivability\,\cite{dustimp2}, to play a pivotal role in the formation of dust accumulation sites\,\cite{introd10,dustimp3} and to constitute a source of plasma-facing component (PFC) damage\,\cite{dustimp4}.

Experimental and theoretical investigations mainly focused on the low-to-moderate impact speed range, $v_{\mathrm{imp}}\lesssim200$\,m/s, being the most typical for micrometer dust-wall collisions in tokamaks\,\cite{impactI1,impactI2,impactI3,impactI4}. Studies also targeted the hyper-velocity range, $v_{\mathrm{imp}}\gtrsim4000$\,m/s, given the possibility for excessive wall damage, since it is characterized by complete dust vaporization, deep crater formation and fast secondary solid ejecta production\,\cite{impactI5,impactI6,impactI7,impactI8,impactI9,impactI0}. However, hyper-velocity dust speeds are currently deemed as rather unrealistic in tokamaks. On the other hand, no investigation has concentrated on the high velocity range, $200\lesssim{v}_{\mathrm{imp}}[\mathrm{m/s}]\lesssim4000$, that is characterized by strong deformation of dust and shallow crater formation\,\cite{impactII}. It has been recently concluded that the high velocity range is attainable for tokamak dust\cite{dustimp4}. Thus, high velocity impacts comprise an unexplored source of wall erosion and an unidentified mechanism of tokamak-born dust destruction.

The above naturally bring forth the question of whether high velocity tungsten dust impacts on tungsten PFCs (W-on-W) can be harmful for future fusion reactors. The first step towards an answer requires the formulation of a reliable damage law that correlates the crater volume with the dust size, impact speed and impact angle. In this work, given the lack of computational tools that account for the complex processes that unfold inside the impact site, an experimental study is performed. Nearly monodisperse W spherical dust populations are prepared that are accelerated in a controlled manner to high speeds ($500-3200\,$m/s) towards a W plate with a light gas gun shooting system. Crater characteristics (diameter, depth) are then obtained by means of a scanning electron microscope, an optical microscope and a mechanical profiler. Finally, the large normal impact statistics are exploited for the formulation of reliable scaling laws for the crater characteristics as functions of the impact speed and dust size, which can be ultimately employed for erosion estimates in future reactors.

\section{Aspects of fusion relevant dust-wall impacts}\label{sec:theory}

\noindent Mechanical collisions between solid spherical micrometric projectiles and semi-infinite bulk solid targets have been extensively studied in the literature due to their relevance in geological phenomena, space applications, weapons research and technological applications. For refractory metal projectiles and targets (W, Mo), three impact speed ranges can be distinguished that are further separated in multiple regimes\,\cite{impactII}. In what follows, we shall present the elementary characteristics of each velocity range and briefly discuss its relevance to dust in tokamaks. It is worth pointing out that each velocity range limit depends not only on the material composition but also on the dust size and dust \& wall temperature. Thus, the limits stated below should be deemed as indicative. The discussion is also applicable to oblique impacts, but not to molten projectiles or targets.

In the \textbf{low-to-moderate velocity range}, $v_{\mathrm{imp}}\lesssim200$ m/s, the mechanical impact induces weak plastic deformations to the projectile and the target\,\cite{impact01}. Two regimes can be distinguished; the \emph{sticking regime} of $v_{\mathrm{imp}}\lesssim5\,$m/s where the irreversible adhesive work performed during the near-elastic impact dissipates the incident kinetic energy of the projectile and thus leads to immobilization\,\cite{impact02}, the \emph{inelastic rebound regime} $5\lesssim{v}_{\mathrm{imp}}[\mathrm{m/s}]\lesssim200$ where the projectile rebounds with a decreased kinetic energy due to adhesive work, plastic dissipation and frictional losses\,\cite{impact03}. This is the most typical range for tokamak dust that has been studied experimentally and theoretically for fusion-relevant materials\,\cite{impactI1,impactI2,impactI3,impactI4}. In particular, a unified analytical description has been achieved for both regimes by combining an elastic-perfectly plastic adhesive impact mechanics model\,\cite{impact04} with a rigid sliding body model\,\cite{impact05}.

In the \textbf{high velocity range}, $200\lesssim{v}_{\mathrm{imp}}[\mathrm{m/s}]\lesssim4000$, the mechanical impact induces strong plastic deformations, surface melting and partial fragmentation to both the projectile and target\,\cite{impact06}. Three regimes can be distinguished; the \emph{deformation regime} of $200\lesssim{v}_{\mathrm{imp}}[\mathrm{m/s}]\lesssim500$ that is accompanied by severe projectile flattening \& shallow target crater formation\,\cite{impact07}, the \emph{bonding or cold spray regime} of $500\lesssim{v}_{\mathrm{imp}}[\mathrm{m/s}]\lesssim1000$ that is characterized by the adhesion of the projectile on the target, which can be realized by various mechanisms such as localized melting, adiabatic shear instability, viscous interlocking and interfacial amorphization\,\cite{impact08,impact09,impact10,impact11,impact12}, the \emph{partial disintegration regime} of $1000\lesssim{v}_{\mathrm{imp}}[\mathrm{m/s}]\lesssim4000$ that is accompanied by material splash ejection and partial fragmentation\,\cite{impact06}. It has been recently demonstrated that the high velocity range is relevant for the solid dust that is produced by the explosive stopping of runaway electrons on PFCs. In particular, postmortem and \emph{in situ} evidence from FTU (2013 shutdown and 2019 decommisioning), strongly supported by dust transport simulations and by dedicated laboratory tests, revealed that the striking of runaway electrons on the equatorial poloidal limiter led to the generation of very fast solid dust that generated multiple shallow craters upon impact on the adjacent toroidal limiters\,\cite{dustimp4}.

\begin{figure}
\centering
\includegraphics[width = 3.4in]{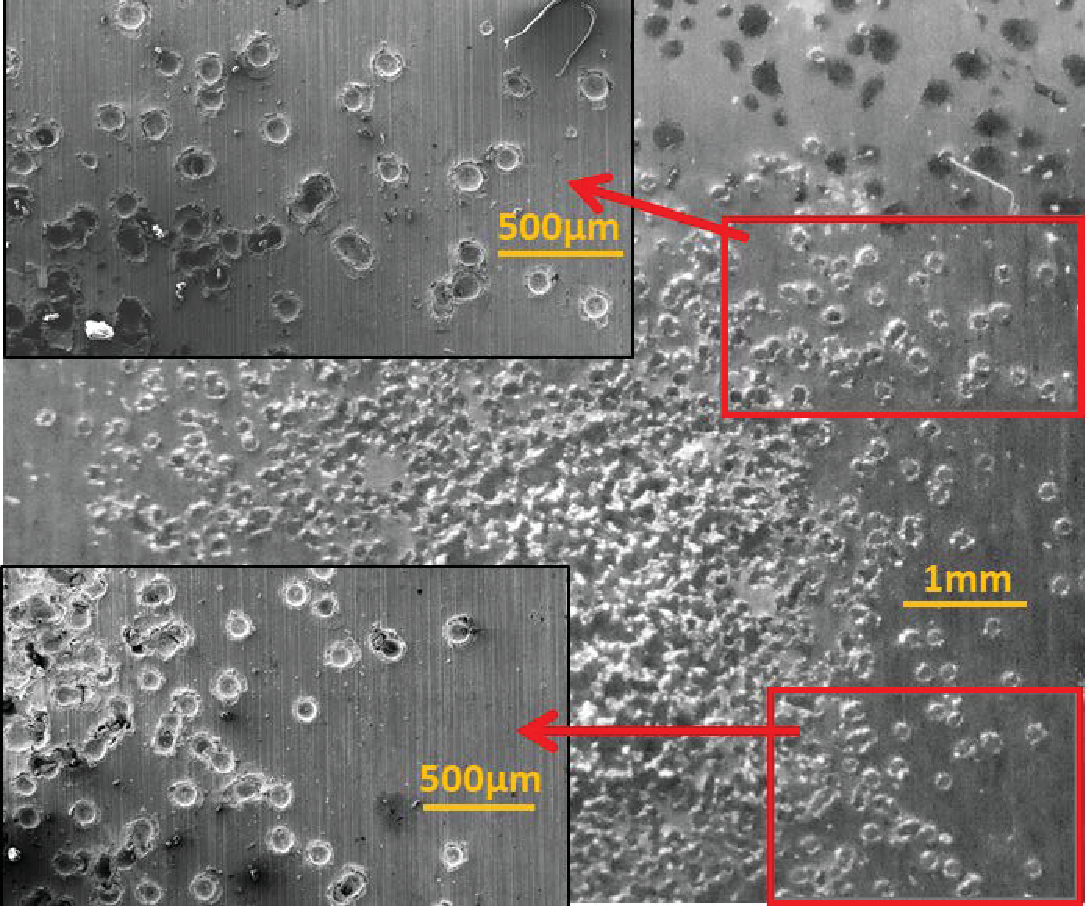}
\caption{Image of a damaged bulk W target after the high velocity impact of spherical W dust (here $D_{\mathrm{d}}=51\,\mu$m and $v_{\mathrm{imp}}=2075\,$m/s) together with magnified SEM images of two regions that have been included in the statistical crater analysis.}\label{fig:statistics-impacts}
\end{figure}

\begin{table}
  \centering
  \caption{W-on-W impacts realized by the one- and two-stage light gas gun. In what follows, $D_{\mathrm{d}}$ denotes the mean dust diameter, $v_{\mathrm{imp}}$ the impact speed, $\#$ the number of craters, $D_{\mathrm{c}}$ the measured crater diameter, $\sigma_{\mathrm{D}}$ the standard deviation in the crater diameter, $H_{\mathrm{c}}$ the measured crater depth, $\sigma_{\mathrm{H}}$ the standard deviation in the crater depth. The negative sign in the crater depth is indicative of impact bonding.}\label{Tab_collection}
\begin{tabular}{c c c c c c c }
$D_{\mathrm{d}}$ & $v_{\mathrm{imp}}$ & $\#$  & $D_{\mathrm{c}}$ & $\sigma_{\mathrm{D}}$ & $H_{\mathrm{c}}$ & $\sigma_{\mathrm{H}}$    \\
($\mu$m)         & (m/s)              &       & ($\mu$m)         & ($\mu$m)              & ($\mu$m)         & ($\mu$m)                 \\ \hline\hline
$51$             & $660$              & $97$  & $74.2$           & $15.5$                & $+7.5$           & $3.3$                    \\
$51$             & $984$              & $83$  & $73.0$           & $15.3$                & $-13.3$          & $3.9$                    \\
$51$             & $1565$             & $62$  & $84.0$           & $17.7$                & $+14.9$          & $5.7$                    \\
$51$             & $2075$             & $73$  & $99.3$           & $20.9$                & $+32.2$          & $5.1$                    \\
$51$             & $2506$             & $47$  & $106.1$          & $22.3$                & $+40.4$          & $5.3$                    \\
$51$             & $2561$             & $36$  & $107.8$          & $22.9$                & $+38.4$          & $7.1$                    \\
$51$             & $3128$             & $37$  & $115.7$          & $24.4$                & $+50.0$          & $6.7$                    \\ \hline
$63$             & $596$              & $110$ & $84.2$           & $17.5$                & $+8.1$           & $3.2$                    \\
$63$             & $764$              & $64$  & $89.0$           & $18.4$                & $-26.3$          & $4.2$                    \\
$63$             & $1012$             & $63$  & $89.2$           & $18.2$                & $-14.3$          & $4.6$                    \\
$63$             & $1534$             & $72$  & $97.8$           & $20.5$                & $+21.4$          & $6.4$                    \\
$63$             & $2039$             & $78$  & $119.0$          & $24.8$                & $+43.3$          & $7.3$                    \\
$63$             & $2485$             & $37$  & $129.3$          & $26.8$                & $+51.1$          & $8.1$                    \\
$63$             & $2500$             & $45$  & $130.9$          & $27.4$                & $+47.5$          & $8.2$                    \\
$63$             & $3108$             & $33$  & $142.0$          & $35.1$                & $+62.4$          & $11.3$                   \\
$63$             & $3190$             & $36$  & $149.3$          & $30.8$                & $+66.6$          & $8.2$                    \\ \hline
$76$             & $583$              & $99$  & $104.9$          & $21.3$                & $+8.7$           & $3.2$                    \\
$76$             & $997$              & $48$  & $108.2$          & $22.2$                & $-16.8$          & $5.8$                    \\
$76$             & $1563$             & $57$  & $121.8$          & $26.6$                & $+31.6$          & $10.5$                   \\
$76$             & $2033$             & $35$  & $143.3$          & $29.4$                & $+50.1$          & $7.3$                    \\
$76$             & $2058$             & $43$  & $145.6$          & $30.7$                & $+48.1$          & $5.8$                    \\
$76$             & $2513$             & $25$  & $153.8$          & $31.4$                & $+64.4$          & $8.9$                    \\
$76$             & $2551$             & $36$  & $152.6$          & $33.4$                & $+65.3$          & $10.7$                   \\
$76$             & $3069$             & $27$  & $180.7$          & $42.3$                & $+81.9$          & $14.2$                   \\
$76$             & $3126$             & $20$  & $180.0$          & $38.4$                & $+80.1$          & $9.2$                    \\ \hline\hline
\end{tabular}
\end{table}

\begin{figure*}[!h]
\centering
\includegraphics[width = 5.90in]{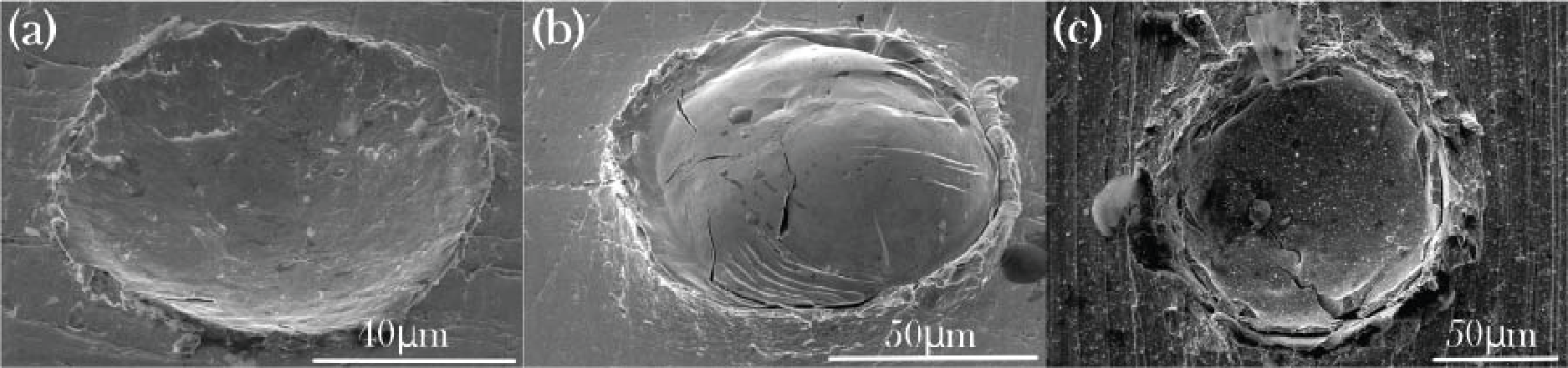}
\caption{SEM images of the three erosion regimes that are realized in high velocity W-on-W normal dust-wall impacts for a given spherical dust size ($D_{\mathrm{d}}=63\,\mu$m): (a) \textbf{plastic deformation regime} (here for $v_{\mathrm{imp}}=596\,$m/s), tilted image at $45^{\circ}$, (b) \textbf{impact bonding regime} (here for $v_{\mathrm{imp}}=764\,$m/s), tilted image at $45^{\circ}$, (c) \textbf{partial disintegration regime} (here for $v_{\mathrm{imp}}=1563\,$m/s), normal image at $90^{\circ}$. Note the subtle morphological differences between (a) and (c); in the partial disintegration regime, the crater rim is rougher as well as more elevated and small fragments or melt splashes are present within the crater cone.}\label{fig:regimes-impacts}
\end{figure*}

In the \textbf{hyper-velocity range} of ${v}_{\mathrm{imp}}\gtrsim4000\,$m/s, the impact velocity is comparable or larger than the compressive sound speeds of the projectile and the target\,\cite{impactII}. This triggers the emergence of strong shockwaves, upon whose sudden compaction extreme transient pressures and temperatures arise at the collision zone. Upon shockwave release, both materials extensively vaporize while the localized energy is large enough to cause atomic excitation and partial ionization\,\cite{impact13}. As a result, hyper-velocity impacts are accompanied by strong electrostatic\,\cite{impact13,impact14} and spectroscopic signals\,\cite{impact15,impact16}. In addition, the emerging craters feature excavated volumes that by far exceed the volume of the projectile. Finally, very high velocity solid ejecta can be released that cause further damage to the surrounding material\,\cite{impact17}. Solid evidence of hypervelocity impacts have been obtained in FTU that concerned the in-situ detection of dust impact ionization by electrostatic probes and the post-mortem observation of craters\,\cite{impactI5,impactI6,impactI7}. However, considering the typical initial dust velocity distributions and given the generally restricted acceleration lengths, no acceleration mechanisms are currently known that can lead to impact speeds within the hyper-velocity range\,\cite{introdu4,introdu8}.

\section{Experimental}\label{sec:experiment}

\noindent High-sphericity low internal porosity W dust was purchased from \enquote{TEKNA Plasma Systems}. The original batch had a nominal size distribution of $45-90\,\mu$m. From this polydisperse batch, three nearly monodisperse sub-populations
were meshed out using a sequence of six sieves with nominal sizes of $80,\,75,\,71,\,63,\,56$ and $50\,\mu$m. The mean W dust diameters are $51(\pm5)\mu$m, $63(\pm5)\mu$m and $76(\pm5)\mu$m, sizes that are comparable to the most probable sizes of the fast Mo solid dust that was observed in FTU\,\cite{dustimp4}. The bulk square W targets have $23\,$mm length and $4\,$mm thickness.

High velocity dust-wall impacts are realized by means of a light gas gun system\,\cite{lightga1,lightga2}. In the \emph{two stage configuration}, the first stage is a high pressure reservoir connected to the second stage with a fast valve. The second stage (or pump tube) is a cylinder in which the light gas is fed at relatively low pressure and compressed by a free piston following the fast valve operation. The compressed light gas rapidly expands into the launch tube, which has a smaller diameter than the pump tube, simultaneously accelerating a macro-scale projectile (sabot) which features a cavity that is loaded with micron dust. The dust particles are separated from the sabot at the end of the launch tube where the latter impinges on a diaphragm. The dust particles free stream with high speeds into a vacuum chamber that features the target. Impact speeds within the range of $1000\lesssim{v}_{\mathrm{imp}}[\mathrm{m/s}]\lesssim3500$ have been realized by employing different light gases (nitrogen or hydrogen) and setting different initial pressures for the first ($20-40\,$bar) \& second ($0.8-1.2\,$bar) stage. In the \emph{single stage configuration}, the high-pressure hydrogen reservoir is directly connected to the launch tube without an intervening piston and the achieved impact speeds lie within $400\lesssim{v}_{\mathrm{imp}}[\mathrm{m/s}]\lesssim1000$.

The dust speed is measured through the dust transit time between two laser sheets with an uncertainty that is generally less than $\pm1\%$. This measurement error primarily originates from the relatively large thickness of the individual beams ($1\,$mm) when compared to the beam spacing ($100\,$mm). The optical method was preferred over the time-of-flight method\,\cite{lightga2,lightga3}, where the start trigger is provided by the muzzle flash due to sabot ejection from the barrel and the stop trigger is provided by the impact flash, whose precision is $\sim\pm3\%$. The dust cloud speed should well approximate the dust particle speed, given the near-constant small cloud width recorded by the laser sheets.

Overall, $34$ impact tests were performed with the three dust sub-populations and impact speeds covering the high velocity range. Light gas gun acceleration leads to a large number of W-on-W impacts, since it is not possible to load a small dust number on the sabot cavity. However, the useful crater statistics are rather restricted, since the central spot features many overlapping craters that should be excluded from the analysis (see figure \ref{fig:statistics-impacts}) and since the central spot also features few visibly smaller craters due to dust-dust impacts in the proximity of the target (see the finite width of the dust cloud). Naturally, it is possible to minimize the central spot by loading less dust on the sabot, but this compromises the accuracy of the speed measurement and does not necessarily improve statistics. From the $34$ shot samples, $25$ featured a large enough number of isolated craters suitable for statistical analysis (Table \ref{Tab_collection}).

The shot W-on-W samples were first mapped by means of a Scanning Electron Microscope (SEM), at low magnification and at high resolution, in order to statistically estimate the average value of the crater diameter at different impact conditions. The crater depth was then measured with a precision optical microscope (Leitz Wetzlar Ortolux) of $0.5\,\mu$m sensitivity. Moreover, few craters from each shot sample were mapped with a mechanical profiler (KLA-Tencor mod.\,P-15) in order to verify the precision of the above techniques. The instrumental uncertainty was estimated to be $\pm3\,\mu$m for the crater depth (optical) and $\pm15\%$ for the crater diameter (SEM). The average values and standard deviations of the crater dimensions (instrumental and statistical errors) are listed in Table \ref{Tab_collection}.

Since the impact speeds varied from $583\,$m/s to $3190\,$m/s, all three impact regimes of the high velocity range were observed for all the three dust sub-populations. Characteristic examples are illustrated in figure \ref{fig:regimes-impacts}.

\begin{figure*}
\centering
\includegraphics[width = 6.15in]{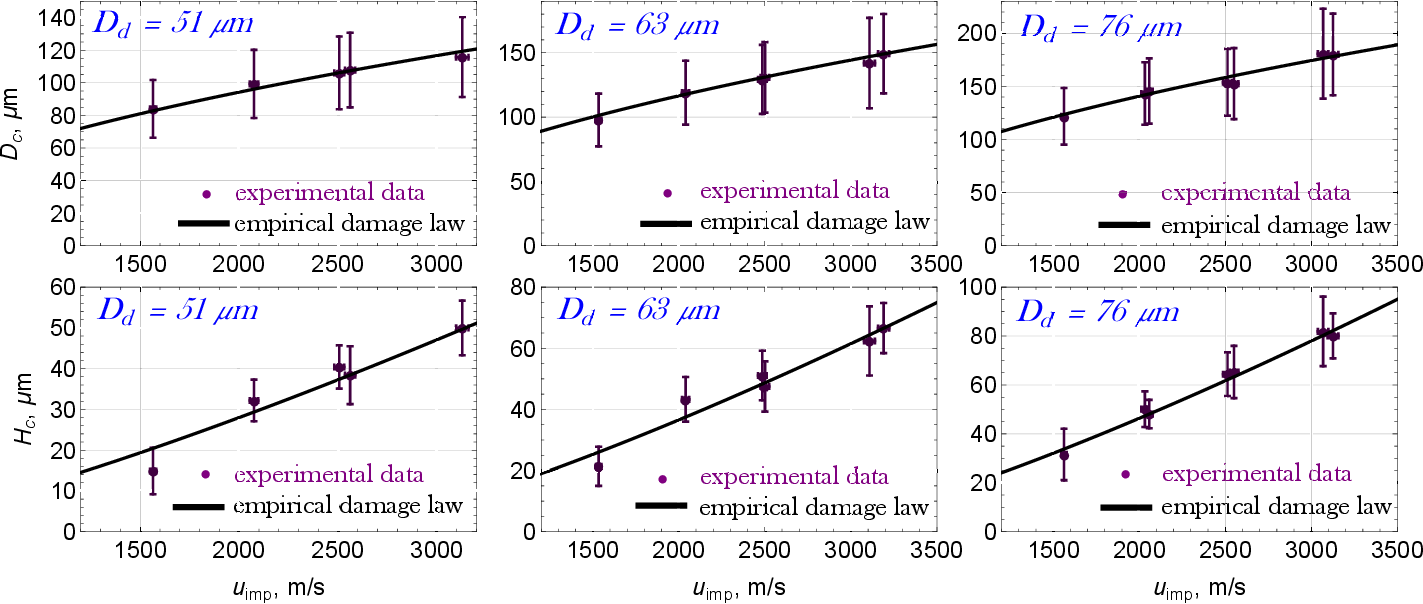}
\caption{The crater diameter (top panel) and crater depth (bottom panel) within the \emph{partial disintegration regime} as functions of the normal W-on-W speed for three spherical W dust sub-populations. Experimental results (purple symbols with error bars) together with the predictions of the empirical damage laws of Eqs.(\ref{ourdamagediameter},\ref{ourdamagedepth}) (solid black lines).}\label{fig:empirical_fits}
\end{figure*}

\section{Analysis}\label{sec:anal}

\subsection{Empirical damage laws}\label{subsec:damage}

\noindent Impact damage laws are empirical correlations of the crater diameter and depth as function of the target and projectile material properties, impact speed, impact angle and dust size\,\cite{hyperve1,hyperve2,hyperve3}. Systematic hyper-velocity dust impact experiments have led to the formulation of empirical damage laws that are applicable in quite extended impact velocity and dust size ranges\,\cite{hyperve1,hyperve2,hyperve3,hyperve4,hyperve5,hyperve6,hyperve7}. Considering their validity for various target-projectile combinations, such general damage laws are of limited accuracy, especially in the high velocity impact range. In what follows, we utilize our extensive W-on-W experimental data to extract empirical damage laws valid within the disintegration regime of the high velocity range, $1\lesssim{v}_{\mathrm{imp}}[\mathrm{km/s}]\lesssim3.5$.

The experimental crater depth and crater diameter data are fitted in the standard power law form of $\alpha({D}_{\mathrm{d}})^{\beta}({v}_{\mathrm{imp}})^{\gamma}$ with $\alpha,\beta,\gamma$ the fitting parameters. The two-variable three-parameter non-linear fit led to the empirical damage laws
\begin{align}
D_{\mathrm{c}}&=0.0330(D_{\mathrm{d}})^{1.005}(v_{\mathrm{imp}})^{0.527}\,,\label{ourdamagediameter}\\
H_{\mathrm{c}}&=0.0000114(D_{\mathrm{d}})^{1.264}(v_{\mathrm{imp}})^{1.282}\,,\label{ourdamagedepth}
\end{align}
with $D_{\mathrm{c}},\,H_{\mathrm{c}},\,D_{\mathrm{d}}$ measured in $\mu$m and $v_{\mathrm{imp}}$ in m/s. The mean absolute relative fitting error is $1.87\%$ for the diameter and $6.89\%$ for the depth. The standard errors for the fitting parameters of the diameter are $0.0089\,(\alpha)$,\,$0.041\,(\beta)$, $0.027\,(\gamma)$. The standard errors for the fitting parameters of the depth are $0.00000803\,(\alpha)$, $0.098\,(\beta)$ and $0.072\,(\gamma)$. The empirical damage laws are plotted in figure \ref{fig:empirical_fits}.

\begin{figure}
\centering
\includegraphics[width = 2.8in]{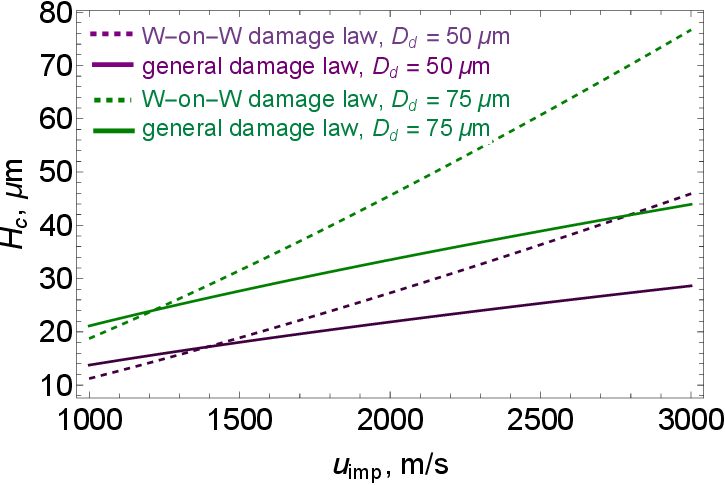}
\caption{The crater depth in the \emph{partial disintegration regime} of the high velocity range as a function of the W-on-W impact speed for two dust sizes. Comparison of the predictions of our specific damage law, Eq.(\ref{ourdamagedepth}), with the general damage law, Eq.(\ref{generaldamagedepth}).}\label{fig:damage-comparison}
\end{figure}

An established general damage law for the crater depth, appropriate for metallic dust exceeding $50\,\mu$m and impact speeds within $2\lesssim{v}[\mathrm{km/s}]\lesssim12$, reads as\,\cite{hyperve1,hyperve2,hyperve3}
\begin{equation}
H_{\mathrm{c}}=5.24\frac{D_{\mathrm{d}}^{19/18}}{h_{\mathrm{t}}^{1/4}}\left(\frac{\rho_{\mathrm{d}}}{\rho_{\mathrm{t}}}\right)^{1/2}\left(\frac{v_{\mathrm{imp}}}{c_{\mathrm{t}}}\right)^{2/3}\,,\label{generaldamagedepth}
\end{equation}
where $H_{\mathrm{c}}$ is the crater depth in cm, $D_{\mathrm{d}}$ the dust diameter in cm, $h_{\mathrm{t}}$ the target's Brinell hardness number, $\rho_{\mathrm{d}},\rho_{\mathrm{t}}$ the mass densities of the dust and the target, $c_{\mathrm{t}}$ the target's sound speed and $v_{\mathrm{imp}}$ the dust impact speed. It is evident from the exponents that the general damage law exhibits a weaker dependence on both the dust size and the impact speed than the W-on-W damage law. More specifically, as discerned from figure \ref{fig:damage-comparison}, the general damage law significantly\,underestimates the crater depth for large dust sizes and high impact speeds. This confirms the importance of target-projectile specific damage laws as those of Eqs.(\ref{ourdamagediameter},\ref{ourdamagedepth}).

\begin{figure}
\centering
\includegraphics[width = 2.9in]{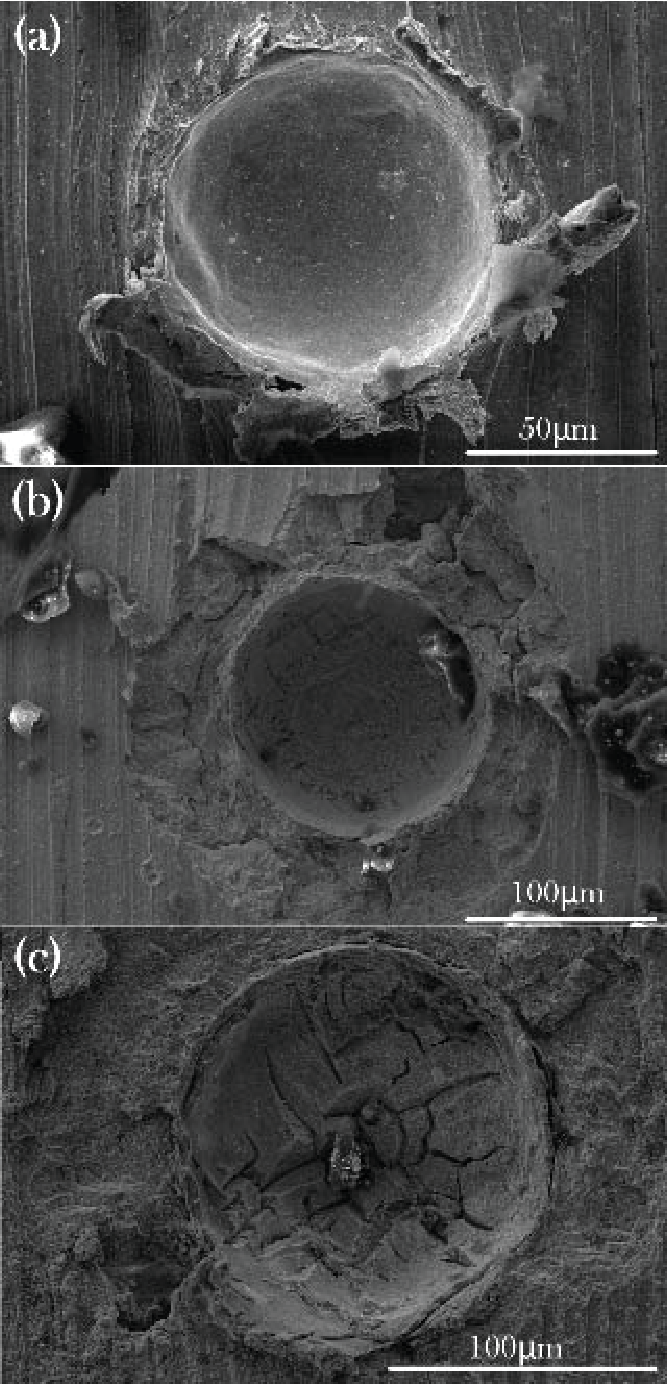}
\caption{SEM images of high velocity W-on-W normal dust impacts that fall into the \emph{partial disintegration regime} for different dust sizes and different impact speeds: (a) $D_{\mathrm{d}}=51\,\mu$m, $v_{\mathrm{imp}}=1506\,$m/s, (b) $D_{\mathrm{d}}=63\,\mu$m, $v_{\mathrm{imp}}=3190\,$m/s, (c) $D_{\mathrm{d}}=76\,\mu$m, $v_{\mathrm{imp}}=2033\,$m/s.}\label{fig:HV-Collisions}
\end{figure}

\begin{figure}
\centering
\includegraphics[width = 2.9in]{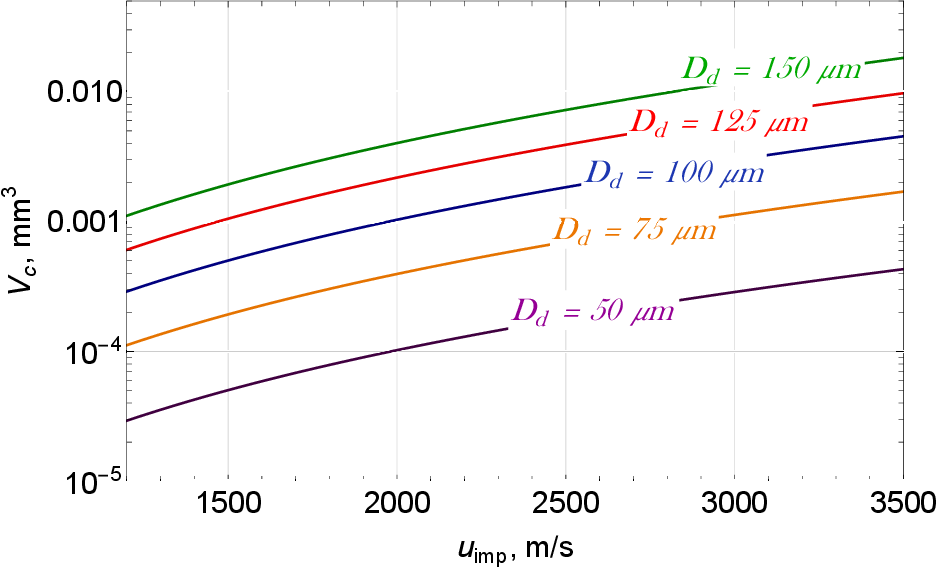}
\caption{The excavated volume predictions of the empirical damage laws for the crater diameter and the crater depth in the \emph{partial disintegration regime} under the spherical cap assumption. Excavated volume as a function of the impact speed for various W dust sizes.}\label{fig:excavated_volume}
\end{figure}

Finally, empirical damage laws allow for an estimate of the excavated volume provided that an additional assumption is made for the geometry of the crater. As expected from symmetry considerations arising from the high dust sphericity, local target planarity and normal nature of the impact, the crater geometry can be well-approximated by a spherical cap. This has been consistently observed in the laboratory impact tests, see figure \ref{fig:HV-Collisions}, and is consistent with the experimental hyper-velocity literature\,\cite{hyperve8,hyperve9}. Thus, the excavated volume can be approximated by
\begin{equation}
V_{\mathrm{c}}=\frac{1}{6}\pi{H}_{\mathrm{c}}\left(\frac{3}{4}D_{\mathrm{c}}^2+H_{\mathrm{c}}^2\right)\,.\label{ourdamagevolume}
\end{equation}
Some characteristic examples are provided in figure \ref{fig:excavated_volume}. Note that the excavated material from a single impact crater is much larger than the excavated material from a single unipolar arc crater\,\cite{arcingC1,arcingC2}, but much smaller than the material excavated during an edge-localized mode or major disruption driven melt event\,\cite{arcingC3,arcingC4}.

\subsection{Critical velocities}\label{subsec:critical}

\noindent Here, we shall discuss the transition between the high velocity regimes for normal W-on-W impacts, which dictates the applicability limits of our empirical damage laws.

The \emph{critical erosion velocity} $v_{\mathrm{crit}}^{\mathrm{ero}}$ is the transition velocity between the bonding and disintegration regimes of the high velocity range. An analytical expression has been derived for $v_{\mathrm{crit}}^{\mathrm{ero}}$ that associates the transition with the impact kinetic energy that triggers melting\,\cite{impact06}. A first-order energy balance analysis that contains a number of approximations leads to the expression\,\cite{impact06}
\begin{equation*}
v_{\mathrm{crit}}^{\mathrm{ero}}=\left(\frac{20e_{\mathrm{th}}I_{\mathrm{melt}}}{\rho_{\mathrm{d}}\sqrt{D_{\mathrm{d}}}}\right)^{2/5}\,,\label{criticalerosion}
\end{equation*}
where $e_{\mathrm{th}}=\sqrt{\rho_{\mathrm{d}}k_{\mathrm{d}}c_{\mathrm{pd}}}$ is the so-called thermal effusivity with $k_{\mathrm{d}}$ the thermal conductivity and $c_{\mathrm{pd}}$ the specific isobaric heat capacity, where $I_{\mathrm{melt}}=T_{\mathrm{m}}-T_0+\Delta{h}_{\mathrm{f}}/c_{\mathrm{pd}}$ is a melting index with $T_{\mathrm{m}}=3695\,$K the melting point of W, $T_0=300\,$K the room temperature and $\Delta{h}_{\mathrm{f}}$ the latent heat of fusion. This expression should be expected to underestimate the critical erosion velocity but has exhibited a good agreement with dedicated Sn, Bi, Zn, Ti experiments\,\cite{impact06}.

The \emph{critical bonding velocity} $v_{\mathrm{crit}}^{\mathrm{bond}}$ is the transition velocity between the deformation and bonding regimes of the high velocity range. A closed-form correlation is available for $v_{\mathrm{crit}}^{\mathrm{bond}}$ that is based on numerical simulations within the assumption that impact bonding can be attributed solely to the onset of adiabatic shear instabilities\,\cite{impact08}. Additional numerical modelling and experimental results allowed the consideration of size effects in the original expression\,\cite{impact09}. The final expression reads as\,\cite{impact08,impact09}
\begin{equation*}
v_{\mathrm{crit}}^{\mathrm{bond}}=\left[667-0.014\rho_{\mathrm{d}}+0.08(T_{\mathrm{m}}-T_0)+10^{-7}\sigma_{\mathrm{y}}\right](D_{\mathrm{d}})^{-0.07}\,,\label{criticalbonding}
\end{equation*}
where all quantities are in SI units and $\sigma_{\mathrm{y}}$ denotes the yield strength. It is worth pointing out that the adopted size exponent is a factor of two less than its traditional value\,\cite{impact09} in accordance with state-of-the-art experiments\,\cite{impact10}.

\begin{figure}
\centering
\includegraphics[width = 2.7in]{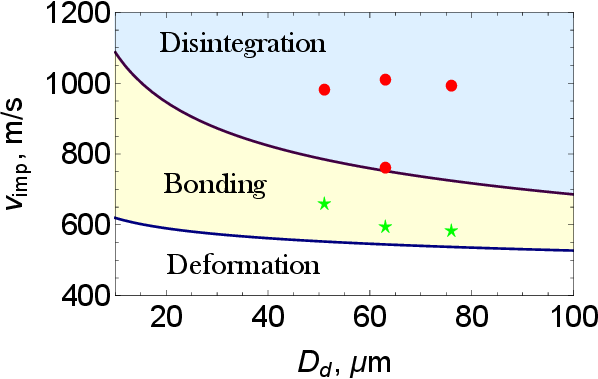}\vspace{-3.0mm}
\caption{Critical erosion (purple line) and bonding velocities (blue line) versus the dust size for normal W-on-W impacts.\,Impacts within the deformation (green stars) and bonding regime (red circles).}\label{fig:critical-velocities}
\end{figure}

In figure \ref{fig:critical-velocities}, the two W-on-W critical velocities are plotted as a function of the dust size. The room temperature recommendations of Ref.\cite{tungsten} have been followed for the W thermophysical properties. Velocity-size combinations that have been experimentally verified to lead to impacts that belong to the deformation and bonding regimes have also been included. It is evident that the aforementioned general expressions underestimate both critical velocities.

\section{Summary and future work}

\noindent The first experimental study of high velocity W dust - W wall impacts, relevant for cascade-like PFC damage caused by runaway electron stopping on PFCs, has been carried out. Controlled normal high velocity impacts have been realized with a light gas dust gun. Surface analysis allowed the identification of three wall damage regimes (plastic deformation, impact bonding, partial disintegration) and the extraction of empirical damage laws for the most harmful disintegration regime. Work in progress focuses on the extension of the impact data to obtain more accurate damage laws, the determination of the critical velocities that demarcate the damage regimes and the realization of oblique impacts. In the future, we aim to quantify the effect of elevated dust temperatures; a step that is particularly important given that explosive runaway electron termination on PFCs is expected to produce fast hot solid dust.

Concerning the potential significance of high velocity solid dust impacts in fusion devices, the following remark is important. While primary runaway electron induced damage could have a major extent, it has a localized character and it could be constrained to sacrificial limiters or to replaceable divertor plates in future fusion reactors. On the other hand, secondary high velocity dust impact damage should be considerably less dramatic, but it has a delocalized character and could thus compromise the integrity of plasma facing components including diagnostics.

\section*{Acknowledgments}

\noindent The work has been performed within the framework of the EUROfusion Consortium,\,funded by the European Union via the Euratom Research and Training Programme (Grant Agreement No\,101052200 - EUROfusion). Views and opinions expressed are however those of the authors only and do not necessarily reflect those of the European Union or European Commission.\,Neither the European Union nor the European Commission can be held responsible for them.

\end{document}